\documentclass[a4paper]{jpconf}
\usepackage{graphicx}

  \newcommand {\nc} {\newcommand}
  \nc {\beq} {\begin{eqnarray}}
  \nc {\eeq} {\nonumber \end{eqnarray}}
  \nc {\eeqn}[1] {\label {#1} \end{eqnarray}}
  \nc {\eol} {\nonumber \\}
  \nc {\eoln}[1] {\label {#1} \\}
  \nc {\ve} [1] {\mbox{\boldmath $#1$}}
  \nc {\ves} [1] {\mbox{\boldmath ${\scriptstyle #1}$}}
  \nc {\mrm} [1] {\mathrm{#1}}
  \nc {\half} {\mbox{$\frac{1}{2}$}}
  \nc {\thal} {\mbox{$\frac{3}{2}$}}
  \nc {\fial} {\mbox{$\frac{5}{2}$}}
  \nc {\la} {\mbox{$\langle$}}
  \nc {\ra} {\mbox{$\rangle$}}
  \nc {\eq} [1] {(\ref{#1})}
  \nc {\Eq} [1] {Eq.~(\ref{#1})}
  \nc {\Ref} [1] {Ref.~\cite{#1}}
  \nc {\Refc} [2] {Refs.~\cite[#1]{#2}}
  \nc {\Sec} [1] {Sec.~\ref{#1}}
  \nc {\chap} [1] {Chapter~\ref{#1}}
  \nc {\anx} [1] {Appendix~\ref{#1}}
  \nc {\tbl} [1] {Table~\ref{#1}}
  \nc {\Fig} [1] {Fig.~\ref{#1}}
  \nc {\ex} [1] {$^{#1}$}
  \nc {\Sch} {Schr\"odinger }
  \nc {\flim} [2] {\mathop{\longrightarrow}\limits_{{#1}\rightarrow{#2}}}
  \nc {\IR} [1]{\textcolor{red}{#1}}
  \nc {\IB} [1]{\textcolor{blue}{#1}}
  \nc{\pderiv}[2]{\cfrac{\partial #1}{\partial #2}}
  \nc{\deriv}[2]{\cfrac{d#1}{d#2}}

\begin{document}

\title{Recent advances in the description of reactions involving exotic nuclei}

\author{Pierre Capel}

\address{Institut f\"ur Kernphysik, Johannes Gutenberg-Universit\"at Mainz, Johann-Joachim-Becher Weg 45, D-55099 Mainz}

\ead{pcapel@uni-mainz.de}

\begin{abstract}
In this contribution to the proceedings of the International Nuclear Physics Conference 2019, I review recent developments made in reaction models used to analyse data measured at radioactive-ion beam facilities to study exotic nuclear structures.
I focus in particular on reactions like elastic scattering and breakup, which are used to study halo nuclei.
Although these peculiar nuclei challenge usual nuclear-structure models, some can now be computed \emph{ab initio}.
This brief review illustrates the progresses made in nuclear-reaction theory in the last few years to improve the description of the projectile within reaction models.
I dedicate this contribution to the memory of Mahir Hussein, who has significantly contributed to this field and who passed away in May this year.
\end{abstract}

\section{Introduction}
The development of radioactive ion beam (RIB) facilities in the mid 1980s has enabled the experimental exploration of the nuclear chart away from stability.
This technical breakthrough has revealed exotic structures unseen in stable nuclei.
While stable nuclei are usually compact objects whose matter radius follows an $A^{1/3}$ law, some nuclei close to the neutron dripline break that law and are much larger than expected.
This unusual size is due to the weak binding of one or two neutrons observed in these nuclei.
Thanks to this loose binding, the valence neutrons can tunnel far away into the classically forbidden region and therefore exhibit a high probability of presence at a large distance from the other nucleons.
They hence form a sort of diffuse halo surrounding a compact core \cite{HJ87}.

Because they challenge usual nuclear-structure models, halo nuclei are the subject of many experimental and theoretical studies \cite{Tan96}.
Being located close to the dripline, halo nuclei are short-lived and therefore cannot be investigated through usual spectroscopic methods.
Information about their structure has to be gathered from indirect techniques.
Reactions performed at RIB facilities, like elastic scattering or breakup, are often used to study the structure of these exotic nuclei \cite{Tan96}.
In order to infer reliable information from such measurements, the reaction mechanism must
be well understood.
An accurate model of the reaction coupled to a realistic description of the projectile is thus needed \cite{BC12}.
Since the early days of RIBs, various models of reactions have been developed, which have helped us better grasp the dynamics of these reactions \cite{BC12}.

Recently particular efforts have been put to improve the description of the exotic projectiles within reaction models (see, e.g., Refs. \cite{SNT06,ML12,DH13,DAL14,Lay16,DCM17,Moro19c}).
Interestingly, \emph{ab initio} models of nuclei are now able to describe halo nuclei \cite{CNR16,Navratil19c} and hence provide nuclear-reaction theorists with inputs to constrain their description of the nuclei within their reaction models.
This can be done \cite{CPH18,MC19} using an effective field theory of halo nuclei \cite{BHK02} (see \Ref{HJP17} for a recent review).

In \Sec{model}, I briefly describe the few-body model of reactions upon which most modern descriptions of reactions are based.
Then, in Secs.~\ref{micro} and \ref{corex}, I present recent developments made to go beyond this few-body description \cite{SNT06,ML12,DH13,DAL14}. 
Finally, in \Sec{EFT}, I explain how an effective-field theory of halo nuclei can be used to constrain the simple few-body model of the reaction based on predictions of \emph{ab initio} calculations \cite{CPH18,MC19}. 
A summary is provided in \Sec{conclusion}.

\section{Few-Body Model of Reactions}\label{model}
\begin{figure}
\begin{minipage}{12pc}
\includegraphics[width=10.5pc]{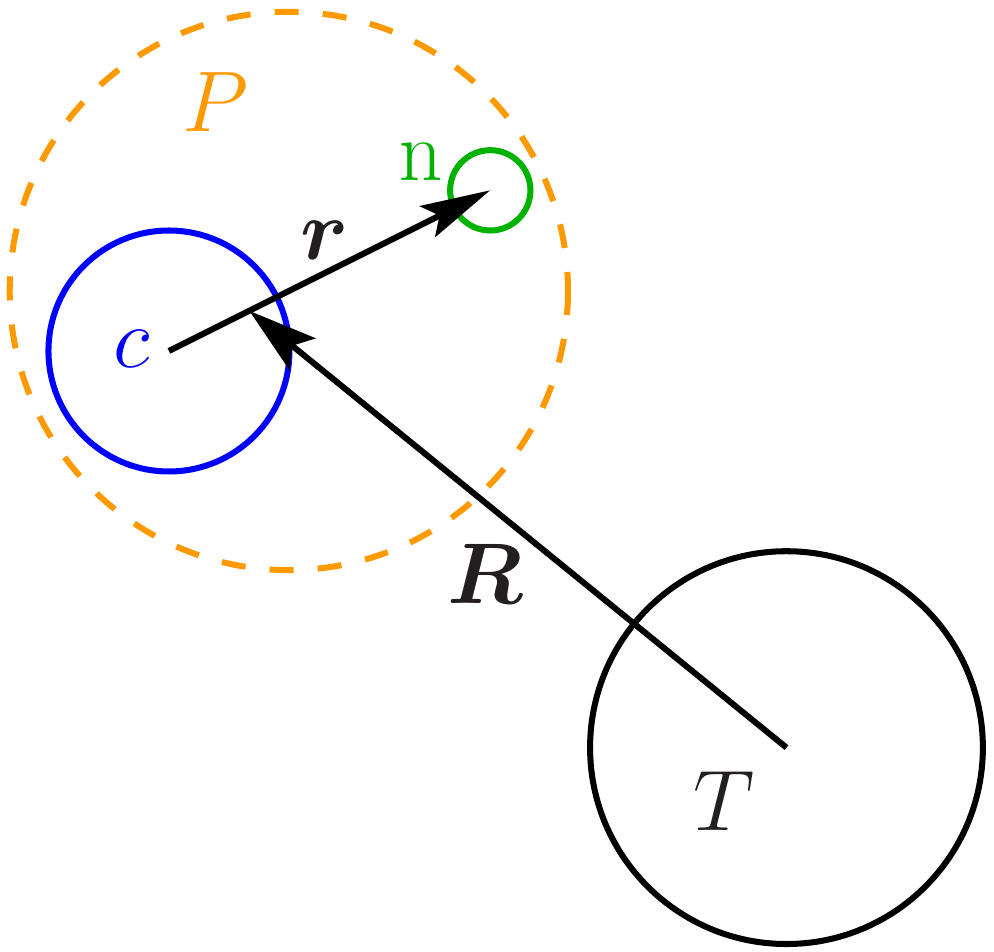}
\end{minipage}
\hspace{2pc}
\begin{minipage}{18pc}\caption{\label{f1}Few-body modelling of a collision of a one-neutron halo nucleus $P$ on a target $T$.
The former is seen as a two-body system: an inert core $c$ to which a neutron $\rm n$ is loosely bound, whereas the structure of the target is usually neglected.
The Jacobi set of coordinates is composed of the $c$-$\rm n$ relative coordinate $\ve{r}$ and the coordinated of the projectile centre of mass to the target $\ve{R}$.}
\end{minipage}
\end{figure}

The usual way to model the collision of a halo nucleus with a target, is based on the few-body framework illustrated in \Fig{f1}.
The projectile $P$ is is described as a two- or three-body quantal system: a core $c$ to which one or two valence neutrons are loosely bound.
The target $T$ is usually seen as a structureless body.

For a one-neutron halo nucleus, the internal structure of the projectile is thus described by the single-particle Hamiltonian
\beq
H_0=-\frac{\hbar^2}{2\mu_{c\rm n}}\Delta_r +V_{c\rm n}(\ve{r}),
\eeqn{e1}
where $\ve{r}$ is the $c$-$\rm n$ relative coordinate and $\mu_{c\rm n}$ is the $c$-$\rm n$ reduced mass.
The potential $V_{c\rm n}$ is an effective interaction that is fitted to reproduce the low-energy spectrum of the projectile: the $c$-$\rm n$ binding energy and the spin and parity of its first states.
The eigenstates of the single-particle Hamiltonian $H_0$ are supposed to describe the $c$-$\rm n$ overlap wave functions obtained from an actual many-body calculation, like the \emph{ab initio} one performed by Calci \etal \cite{CNR16}.

The interaction between the target and the projectile constituents---the core and the halo neutron---are simulated by optical potentials $V_{cT}$ and $V_{{\rm n}T}$, respectively.
The description of the collision between such a two-body projectile and the target hence reduces to solving the following three-body \Sch equation
\beq
\left[-\frac{\hbar^2}{2\mu_{PT}}\Delta_R+H_0+V_{cT}(R_{cT})+V_{{\rm n}T}(R_{{\rm n}T})\right]\Psi(\ve{r},\ve{R})=E_{\rm tot}\, \Psi(\ve{r},\ve{R}),
\eeqn{e2}
where $\ve{R}$ is the relative coordinate of the projectile centre of mass to the target (see \Fig{f1}) and $\mu_{PT}$ is the $P$-$T$ reduced mass.
Equation~\eq{e2} has to be solved with the initial condition that prior to the collision, the projectile is in its ground state $\Phi_0$:
\beq
\Psi(\ve{r},\ve{R})\flim{Z}{-\infty}e^{iKZ}\,\Phi_0(\ve{r})
\eeqn{e3}
where the $Z$ coordinate has been chosen along the beam axis and the wave number $K$ is related to the total energy $E_{\rm tot}$ and the binding energy $\epsilon_0$ of the ground state $\Phi_0$ by $\hbar^2K^2/2\mu_{PT}+\epsilon_0=E_{\rm tot}$.

Various techniques have been proposed to solve that equation (see \Ref{BC12} for a recent review).
In semiclassical approaches, the $P$-$T$ relative coordinate is approximated by a classical trajectory $\ve{R}(t)$ \cite{CBM03c}.
At sufficiently high energy, the eikonal approximation can be made, which simplifies the three-body \Sch equation \eq{e2} by assuming that the $P$-$T$ relative motion does not differ much from the incoming plane wave \eq{e3} \cite{Glauber}.
Equation~\eq{e3} can also be solved fully quantum mechanically within a coupled-channel technique by expanding the three-body wave function $\Psi$ upon the eigenstates of the projectile Hamiltonian $H_0$ \eq{e1}.
Since the breakup channel is the focus of these reaction models, the $c$-$\rm n$ continuum of the projectile has to be included.
This is done by discretising this continuum into energy ``bins'', leading to the Continuum-Discretised Coupled Channel model (CDCC) \cite{Aus87}.

In the following sections, I review recent developments that enable us to go beyond the simple few-body picture presented here and, as such, to get closer to the a microscopic description of reactions involving exotic nuclei, similar to what has been achieved in nuclear structure recently by the advent of \emph{ab initio} models like the one used in Refs.~\cite{CNR16,Navratil19c}.

\section{Microscopic Cluster Model}\label{micro}
The first development I would like to present is the Microscopic CDCC (MCDCC), in which a fully microscopic description of the projectile is included within the CDCC framework \cite{DH13}.
In this work, Descouvemont and Hussein have had the excellent idea to use the microscopic cluster model to describe the projectile as an $A$-body system (see \Ref{DD12} for a recent review).
In that model, the nucleon-nucleon interaction is very simple and the cluster degrees of freedom of the nucleus are assumed a priori, which make this description much simpler compared to the \emph{ab initio} model used in Refs.~\cite{CNR16,Navratil19c}.
Nevertheless, it is a fully microscopic description of the projectile in which the Pauli principle is properly taken into account, which is a significant improvement compared to the simple few-body model described in \Sec{model}.

This microscopic description of the projectile treats both the bound and continuum spectra of the projectile microscopically, which enables to account for the breakup channel in the reaction model.
In addition, it enables building the optical potentials from existing global nucleon-nucleus potentials through a folding technique, hence generating these interactions microscopically.

\begin{figure}
\center
\begin{minipage}{18pc}
\includegraphics[page=4,trim={1.9cm 21.2cm 11cm 1.7cm},clip=true,height=4.6cm]{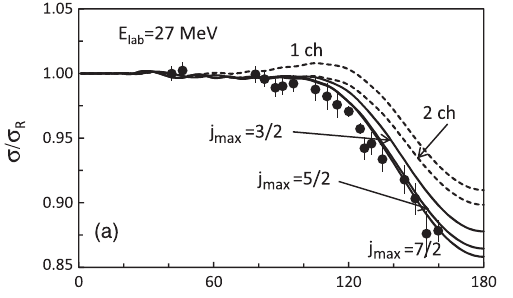}
\end{minipage}
\begin{minipage}{18pc}
\includegraphics[page=6,trim={10.8cm 14.8cm 1.4cm 8.1cm},clip=true,height=4.6cm]{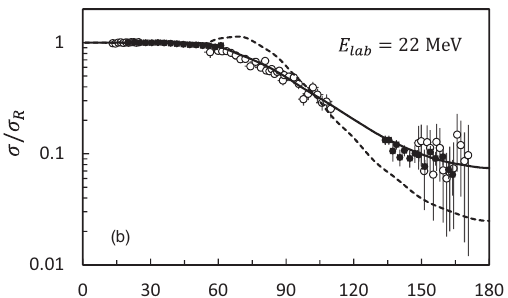}
\end{minipage}
\caption{\label{f2} Elastic-scattering cross section of (left) $^7$Li on Pb at 27~MeV \cite{DH13} and (right) $^6$He on Pb at 22 MeV \cite{Des16}.
The calculations have been performed within the MCDCC \cite{DH13}, which includes a fully microscopic description of the projectile within CDCC.
Reprinted figures with permission from Refs.~\cite{DH13,Des16}. Copyright (2013 and 2016) by the American Physical Society.}
\end{figure}

Figure \ref{f2} illustrates the elastic-scattering cross section obtained within this MCDCC as a ratio to Rutherford for (left) $^{7}$Li impinging on Pb at 27~MeV \cite{DH13} and (right) $^{6}$He on Pb at 22~MeV \cite{Des16}.
While the calculations that consider only the bound states of the projectile in the coupled-channel calculations (dashed lines) miss the data, the inclusion of the breakup channel (solid lines) leads to an excellent agreement with experiment.
The left panel of \Fig{f2} shows that the convergence is reached only after the inclusion of a sufficient number of partial waves in the continuum  \cite{DH13}.

The results shown in \Fig{f2} have been obtained without any fitting parameter, which confirms the validity of this approach and its interest in the analysis of collisions involving loosely bound nuclei.
Unfortunately, this method is, until now, limited to the description of the sole elastic scattering.
Albeit present in the model space, the description of the continuum is not precise enough to enable an accurate extraction of breakup cross sections from the calculations.
Hopefully, this will be improved in a near future, leading to a fully microscopic description of the projectile within a CDCC model of reactions with exotic projectiles.

\section{Accounting for the Excitation of the Core}\label{corex}

Another way to improve the description of the projectile, without resorting to a fully microscopic structure model, is to take into account part of the internal structure of the core of the projectile.
This can be done, for example, assuming a collective model, such as a rigid rotor \cite{NTJ96}. 
This enables including excited states in the core spectrum, which can play a role in the structure of the projectile and/or be excited (or de-excited) during the reaction process.
In such a description, the internal Hamiltonian $H_0$ \eq{e1} becomes
\beq
H_0=H_c(\xi_c)+T_r+V_{c\rm n}(\xi_c,\ve{r}),
\eeqn{e4}
where $H_c$ is the Hamiltonian that describes the internal structure of the core $c$, an which depends on the core internal coordinates $\xi_c$.

An extension of the CDCC framework, coined X-CDCC for eXtended-CDCC, has first been developed by Summers, Nunes and Thompson \cite{SNT06}.
Due to the heavy computational cost of this model, it has not been used much until recently when the Sevilla-Lisbon collaboration has taken up the gauntlet to study various reactions with this improved description of the projectile \cite{DAL14,Lay16,DCM17,Moro19c}.

\begin{figure}
\center
\begin{minipage}{18pc}\includegraphics[page=4,trim={1.5cm 20.1cm 10.7cm 2.4cm},clip=true,height=5.1cm]{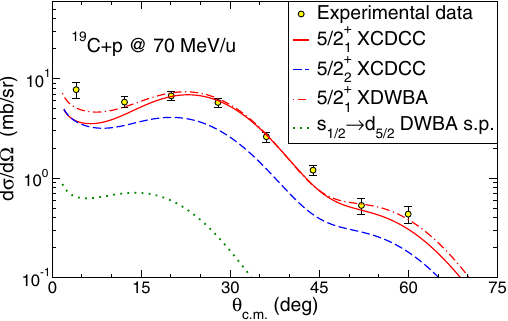}\\
\end{minipage}
\hfill
\begin{minipage}{18pc}
\includegraphics[page=6,trim={2.cm 20cm 11.2cm 2.4cm},clip=true,height=5.2cm]{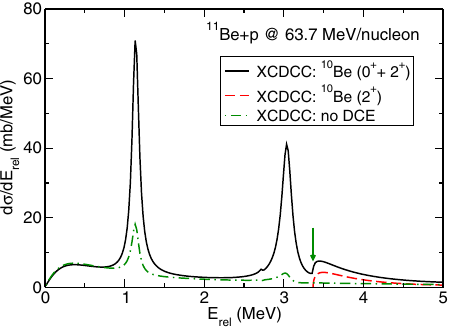}

\vspace{7mm}
\end{minipage}
\caption{\label{f3}Influence of core excitation in the breakup of one-neutron halo nuclei.
Left: $^{19}$C on $\rm p$ at $70A$~MeV \cite{Lay16}. Right: $^{11}$Be on $\rm p$ at $63.7A$~MeV \cite{DCM17}.
Reprinted figures with permission from Refs.~\cite{Lay16,DCM17}. Copyright (2016 and 2017) by the American Physical Society.}
\end{figure}

Results of these studies are illustrated in \Fig{f3}.
The left panel shows the angular distribution following the breakup of $^{19}$C on a proton target at $70A$~MeV restricted to the energy range of its $5/2^+$ resonance in the $^{18}$C-$\rm n$ continuum.
This nucleus has a well known one-neutron halo structure and it is thought that its $^{18}$C core is significantly deformed.
Accordingly, this deformation should be taken into account in the analysis of experiments.
In particular, the $5/2^+$ resonance in the $^{18}$C-$\rm n$ continuum is supposed to be dominated by a configuration in which the core is mostly in its $2^+$ excited state.
This is confirmed by the calculations of Lay \etal \cite{Lay16} shown in the left panel of \Fig{f3}.
Calculations taking into account the core excitation during the reaction reproduce the data nearly perfectly without any parameter fitting (solid and dash-dotted red lines).
A calculation within which the $5/2^+$ resonance is described as a simple $d_{5/2}$ single-neutron resonance, assuming $^{18}$C in its $0^+$ ground state, misses the experimental cross section by an order of magnitude (dotted green line).

The right panel of \Fig{f3} displays the breakup cross section of $^{11}$Be on $\rm p$ at $63.7A$~MeV as a function of the relative energy between the $^{10}$Be core and the neutron after dissociation \cite{DCM17}.
The large peaks observed in this spectrum at $E\simeq1.3$~MeV and 3~MeV correspond to the known $5/2^+$ and $3/2^+$ resonant states of $^{11}$Be, respectively.
The solid black line shows the full XCDCC calculation of this reaction.
The green dash-dotted line corresponds to a calculation in which the dynamical excitation of the core has been switched off.
The significant difference between these two results in the energy range of the resonances confirms that these continuum states are mostly populated through the excitation of the core during the collision.

These results show that core excitation plays a major role in resonant breakup reactions and that interesting information about the structure of the projectile can be obtained from such measurements.
It is therefore needed to include this channel of reaction in the models used to analyse experimental data.

\section{Halo Effective Field Theory}\label{EFT}
In this last section, I present a new idea that enables to easily test predictions of \emph{ab initio} nuclear-structure models while keeping the simple few-body description of the reaction mentioned in \Sec{model}.
The idea is to couple a usual few-body model of the reaction to an effective field theory  description of the halo nucleus, the Halo EFT \cite{BHK02} (see \Ref{HJP17} for a recent review).

Halo EFT is based on the clear separation of scales observed in halo nuclei, which is seen as a compact and tightly bound core to which an extended halo is loosely bound.
Following the usual idea of effective field theory, the Hamiltonian $H_0$ \eq{e1} is then extended upon the small parameter $R_{\rm core}/R_{\rm halo}$, where $R_{\rm core}$ and $R_{\rm halo}$ are, respectively, the sizes of the core and the halo.
In this expansion, the $c$-$\rm n$ interaction is replaced by a contact interaction at leading order (LO) assuming that all short-range physics can be neglected.
For practical use, this contact interaction is regularised by a Gaussian of width $\sigma$, which then stands for the short-range physics neglected in the model.
At higher orders, other terms are added to the interaction, which corresponds to derivatives of the LO Gaussian potential \cite{HJP17}.
At next-to-leading order (NLO) we parametrise the $c$-$\rm n$ interaction as \cite{CPH18}
\beq
V_{c\rm n}^{\rm NLO}(r) = V_0^{lj} e^{-\frac{r ^2}{2 \sigma^2}}+V_2^{lj} r^2e^{-\frac{r ^2}{2 \sigma^2}},
\eeqn{e5}
where $V_0^{lj}$ and $V_2^{lj}$ are parameters fitted to known physical properties of the nucleus in the partial wave $lj$.
Varying $\sigma$ enables us to test the sensitivity of the reaction to the short distances in the projectile description.
At NLO the potential \eq{e5} is fitted up to the $p$ waves, and for higher partial waves the interaction is neglected.

In \Ref{CPH18}, a Halo-EFT description of $^{11}$Be has been included into a precise code of breakup reactions \cite{BCG05} and the results have been compared to experimental data measured at RIKEN \cite{Fuk04}.
In the $s_{1/2}$ and $p_{1/2}$ partial waves, the Halo-EFT potential is fitted to reproduce the binding energy and the asymptotic normalisation coefficient (ANC) of the $1/2^+$ ground state and the $1/2^-$ excited state, respectively.
The binding energy are known experimentally and we use the ANC predicted by the \emph{ab initio} calculations of Calci \etal \cite{CNR16}.
This fit leads also to a good description of the low-energy $s_{1/2}$ and $p_{1/2}$ phase shifts predicted in \Ref{CNR16}.
The same \emph{ab initio} calculation predicts the $p_{3/2}$ phase shift to be negligibly small at low energy.
Accordingly, we consider a nil $c$-$\rm n$ interaction in that partial wave.
This new idea thus enables us to test the quality of these predictions upon experimental data.

\begin{figure}
\center
\begin{minipage}{18pc}
\includegraphics[width=8.5cm]{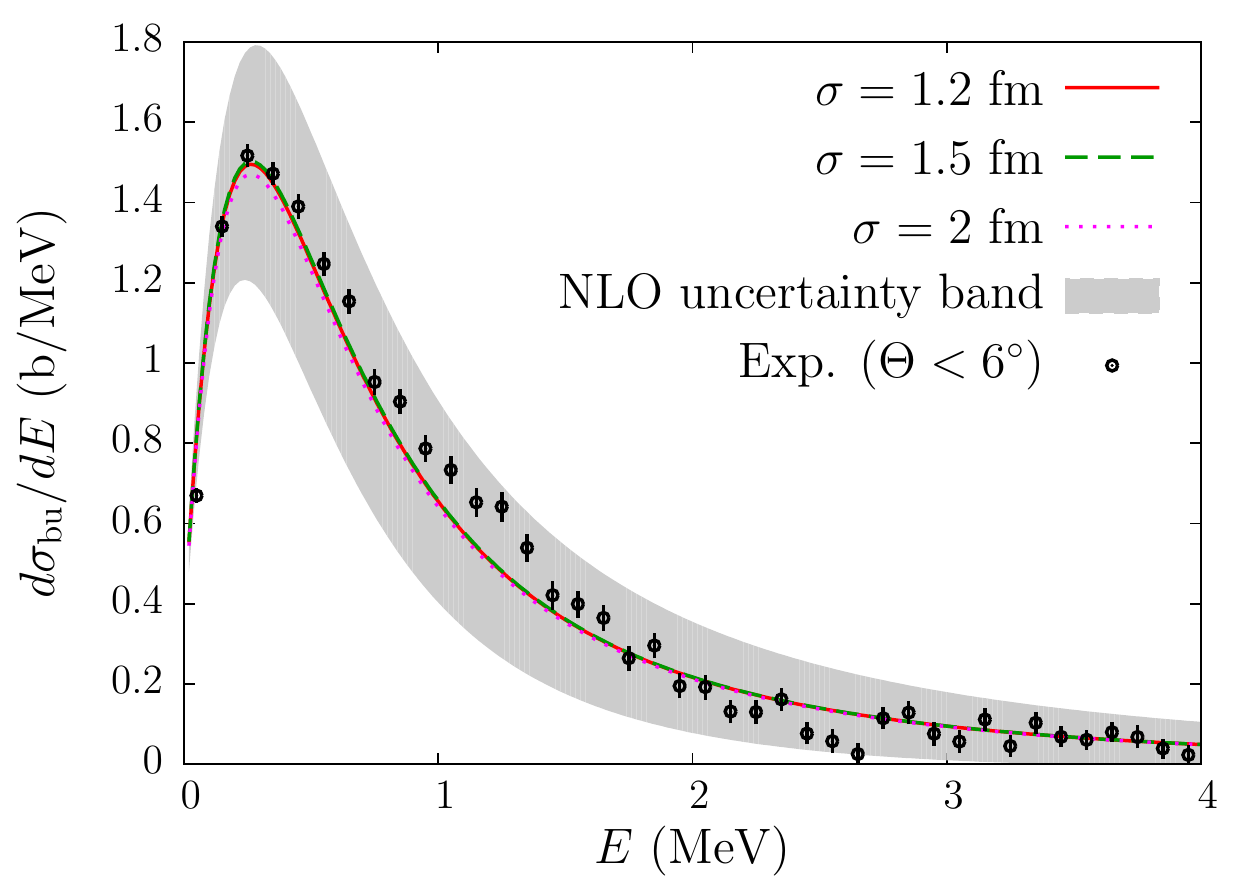}
\end{minipage}
\hfill
\begin{minipage}{18pc}
\includegraphics[width=8.5cm]{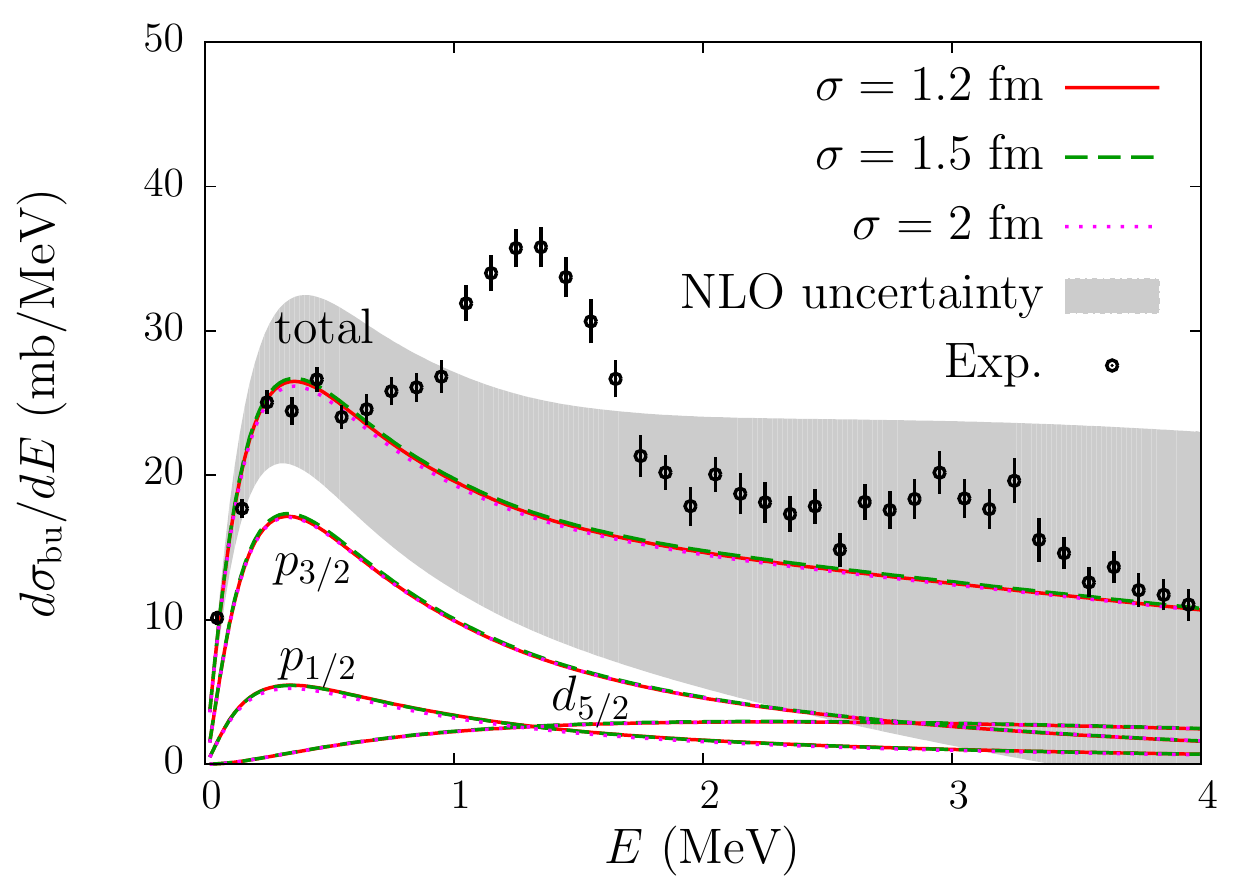}
\end{minipage}
\caption{\label{f4}Including a Halo-EFT description of $^{11}$Be at NLO within a precise model of reaction, we are able to reproduce breakup cross section on (left) Pb and (right) C at about $70A$~MeV \cite{CPH18}.
Data are from \Ref{Fuk04}.
Reprinted figure with permission from \Ref{CPH18}. Copyright (2018) by the American Physical Society.}
\end{figure}

The results of these tests are illustrated in \Fig{f4} for the breakup of $^{11}$Be on (left) Pb at $69A$~MeV and (right) C at $67A$~MeV \cite{Fuk04}.
In both panels, the breakup cross section is plotted as a function of the relative energy $E$ between the $^{10}$Be core and the halo neutron after dissociation.
We observe that on Pb, the agreement with the data is excellent, showing that a simple NLO description of $^{11}$Be is sufficient to reproduce this experiment.
This confirms the validity of the \emph{ab initio} predictions of \Ref{CNR16}.
In addition, we observe no sensitivity at all to the width of the Gaussian potential $\sigma$, showing that the reaction is purely peripheral, in the sense that it probes only the tail of the projectile wave function and not its internal part.
Therefore short-range physics is inaccessible from this reaction observable.
Note that similar results have been obtained at higher energy \cite{MC19,Moschini19c}.

On the carbon target [see \Fig{f4}~(right)], the calculation reproduces less well the data: there is a clear missing breakup strength at $E\simeq1.3$~MeV and 3~MeV.
As discussed in \Sec{corex}, these correspond to the low-energy continuum states of $^{11}$Be, which, in a basic shell model, can be seen as $d$ resonances.
Since the $d$ partial waves are not included in the NLO description of $^{11}$Be, it is not surprising that we do not reproduce these structures.
Nevertheless, the general shape of the cross section is well reproduced and, as already observed on Pb, the calculations are not sensitive to $\sigma$, which shows that, albeit nuclear dominated, this reaction is purely peripheral.

\begin{figure}
\begin{minipage}{18pc}
\includegraphics[width=8.5cm]{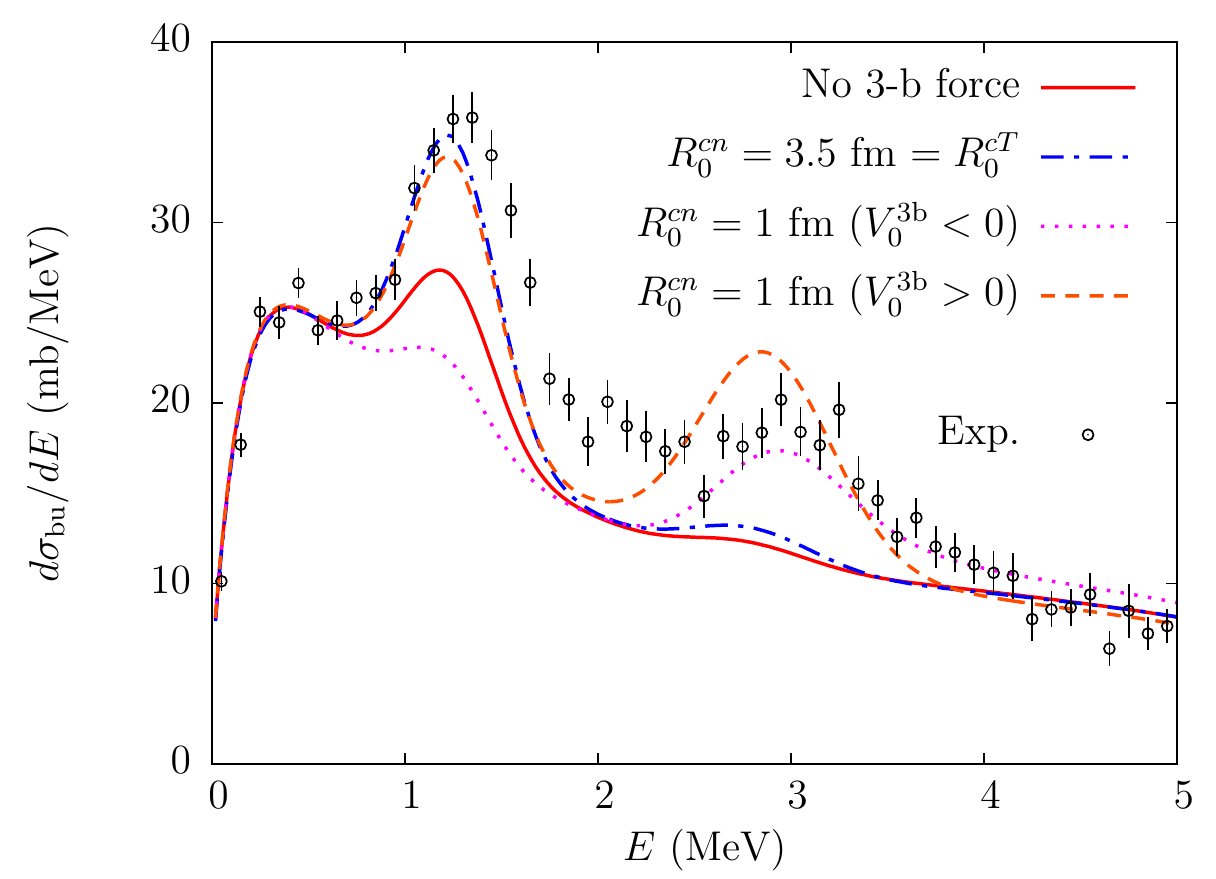}
\end{minipage}
\hspace{2pc}
\begin{minipage}{18pc}\caption{\label{f5}Adding a three-body force between the core, the halo neutron and the target in the calculation of the breakup of $^{11}$Be on C at $67A$~MeV leads to a better description of the experimental data \cite{Fuk04}.
This confirms that dynamical excitation of the core during the collision enhances the resonant breakup of halo nuclei \cite{DCM17,Moro19c}.}
\end{minipage}
\end{figure}

In order to reproduce the structure observed in \Fig{f4}~(right) we need to include the $^{11}$Be resonances.
To do so, we go beyond NLO and include an interaction within the $d_{5/2}$ and $d_{3/2}$ partial waves.
We use the expression \eq{e5} adjusting the parameters of the potential to reproduce the known energy and width of both states.
This leads to the result shown as the solid red line in \Fig{f5}.
As one can see, the presence of a $d_{5/2}$ neutron resonance leads to a clear peak at the energy of the $5/2^+$ state.
The $d_{3/2}$ on the other hand barely affects the calculations.
Following the results of \Ref{DCM17}, we believe that this lack of breakup strength at the energy of the resonances is due to the fact that in our Halo-EFT description of $^{11}$Be we neglect the structure of $^{10}$Be, and in particular the existence of its $2^+$ excited state.
As usual in EFT, the effect of the structure of one of the reactants in a few-body problem can be emulated by a three-body force.
We have devised such an \emph{ad hoc} interaction to simulate the dynamical excitation of the core during the collision \cite{CPH19}.
Depending on the parameters chosen, we can reproduce one (blue dashed-dotted line), the other (magenta dotted line), or both (orange dashed line) resonant peaks in the breakup cross section.
Interestingly, off resonances, the cross section is not affected by the presence of the three-body force.
This confirms the results of Refs.~\cite{DCM17,Moro19c} that the dynamical excitation of the core plays a significant role in resonant breakup.

\section{Summary}\label{conclusion}
Halo nuclei are exotic nuclear structures found at the ridge of the valley of stability.
Their unusually large size poses a challenge to usual nuclear-structure models.
Recently $^{11}$Be has been computed \emph{ab initio} by including the core-neutron degree of freedom within the NCSM \cite{CNR16,Navratil19c}.

Because of their short half-life, these nuclei are probed mostly through indirect techniques, such as reactions.
To reliably infer nuclear-structure information from experimental data, a precise model of the reaction coupled to a realistic description of the projectile is needed.
In this contribution to the proceedings of the International Nuclear Physics Conference, I have presented new ideas in reaction theory that aim at improving this description.

In \Ref{DH13}, Descouvemont and Hussein have introduced a microscopic cluster description \cite{DD12} of the nucleus within a CDCC model of the reaction \cite{Aus87}.
In addition to adding a much more realistic nuclear-model of the projectile, this approach also enables them to deduce the optical potentials that simulate the interaction between the projectile constituents and the target from nucleon-nucleus global potentials.
Excellent agreements with the data are obtained without resorting to any fitting parameter.
Unfortunately, due to the heavy computational cost of this new model, the present results are limited to elastic-scattering calculations \cite{DH13,Des16}.

Still within the CDCC framework, it has been suggested to improve the description of the projectile by considering the core structure in the reaction model by including explicitly some of its low-energy states \cite{SNT06}.
Thanks to this endeavour, it has been shown that the dynamical excitation of the core significantly affects the resonant breakup of halo nuclei \cite{DAL14,Lay16,DCM17,Moro19c}.

Alternatively, it has been suggested to use a Halo-EFT description of the projectile \cite{BHK02,HJP17} within a precise model of reactions \cite{BCG05}.
This provides us with a clean tool to reliably estimate which degrees of freedom are actually probed in the reaction process \cite{CPH18,MC19}.
The Coulomb breakup of $^{11}$Be on Pb, at both intermediate and high energies, is purely peripheral and the experimental data are perfectly reproduced when $^{11}$Be is described at NLO, meaning that only the binding energy and ANC of the bound states and the $p$-wave phase shifts matter in this case.
On C, it is clear that the resonances need to be included to reproduce the data.
This can be done going beyond NLO and adding a $c$-$\rm n$ interaction in the $d$ waves. However, as seen in Refs.~\cite{DCM17,Moro19c}, this single-particle description is not sufficient to reproduce the experimental breakup strength in the energy range of the resonances.
Adding an effective $c$-$\rm n$-$T$ three-body force to the model can emulate the dynamical excitation of the core during the collision and lead to a good agreement with the data \cite{CPH19}.

These few examples illustrate the recent advances made in nuclear-reaction theory, which pave the way to a more precise description of reactions involving exotic nuclei.
These will contribute significantly to a more thorough study of nuclear structure far from stability through reactions.

\section*{Acknowledgments}
This project has received funding from the European Union’s Horizon 2020 research and innovation programme under grant agreement No 654002,
the Deutsche Forschungsgemeinschaft within the Collaborative Research Centers 1044 and 1245, 
and the PRISMA (Precision Physics, Fundamental Interactions and Structure of Matter) Cluster of Excellence.

\section*{References}


\providecommand{\newblock}{}

\end{document}